\documentstyle[12pt,epsf]{article}

\catcode`@=11
\def\chkspace{%
  \relax   
  \begingroup\ifhmode\aftergroup\dochksp@ce\fi\endgroup}
\def\dochksp@ce{%
  \unskip              
  \futurelet\chkspct@k\d@chkspc  
}
\def\d@chkspc{%
  \let\nxtsp@ce=\relax
  \ifx\chkspct@k.\else     
    \ifx\chkspct@k,\else
      \ifx\chkspct@k;\else
        \ifx\chkspct@k!\else
          \ifx\chkspct@k?\else
            \ifx\chkspct@k:\else
              \ifx\chkspct@k)\else
              \ifx\chkspct@k(\else
                \ifx\chkspct@k]\else
                  \ifx\chkspct@k-\else
                    \ifx\chkspct@k\egroup\else  
                      \let\nxtsp@ce=\put@space  
                    \fi
                  \fi
                \fi
              \fi
              \fi
            \fi
          \fi
        \fi
      \fi
    \fi
  \fi
  \nxtsp@ce
}
\def\put@space{$\;$}
\catcode`@=12

\def\ra{\relax\ifmmode \rightarrow\else{{$\rightarrow$}}\fi\chkspace}
\def\Ra{\relax\ifmmode \Rightarrow\else{{$\Rightarrow$}}\fi\chkspace}
\def\etal{{\it et al.}\chkspace}

\def\ie{{\it i.e.}\chkspace}

\def\eg{{\it eg.}\chkspace}

\def\ep{{e$^+$e$^-$}\chkspace}

\def\qu{\relax\ifmmode \quad\else{{$\quad$}}\fi\chkspace}

\def\gluino{\relax\ifmmode \tilde{g} \else $\tilde{g}$ \fi\chkspace}

\def\qq{\relax\ifmmode q\overline{q}
\else $q\overline{q}$ \fi\chkspace}
\def\ff{\relax\ifmmode f\overline{f}
\else $f\overline{f}$ \fi\chkspace}

\def\bb{\relax\ifmmode b\bar{b}
       \else $b\bar{b}$ \fi\chkspace}
\def\cc{\relax\ifmmode { c}\bar{ c}
       \else ${ c}\bar{ c}$ \fi\chkspace}
\def\ccrm{\relax\ifmmode {\rm c}\bar{\rm c}
       \else ${\rm c}\bar{\rm c}$ \fi\chkspace}
\def\tt{\relax\ifmmode {\rm t}\bar{\rm t}
       \else ${\rm t}\bar{\rm t}$ \fi\chkspace}
\def\ss{\relax\ifmmode {\rm s}\bar{\rm s}
       \else ${\rm s}\bar{\rm s}$ \fi\chkspace}
\def\uu{\relax\ifmmode {\rm u}\bar{\rm u}
       \else ${\rm u}\bar{\rm u}$ \fi\chkspace}
\def\dd{\relax\ifmmode {\rm d}\bar{\rm d}
       \else ${\rm d}\bar{\rm d}$ \fi\chkspace}

\def\qqg{\relax\ifmmode q\overline{q}g
\else $q\overline{q}g$ \fi\chkspace}
\def\bbg{\relax\ifmmode b\overline{b}g
\else $b\overline{b}g$ \fi\chkspace}
\def\ccg{\relax\ifmmode c\overline{c}g
\else $c\overline{c}g$ \fi\chkspace}
\def\ttg{\relax\ifmmode t\overline{t}g
\else $t\overline{t}g$ \fi\chkspace}

\def\afb{\relax\ifmmode A_{FB} \else
{{$A_{FB}$}}\fi\chkspace}
\def\afbb{\relax\ifmmode A_{FB}^b \else
{{$A_{FB}^b$}}\fi\chkspace}
\def\pafb{\relax\ifmmode \tilde{A}_{FB} \else
{{$\tilde{A}_{FB}$}}\fi\chkspace}
\def\pafbb{\relax\ifmmode \tilde{A}_{FB}^b \else
{{$\tilde{A}_{FB}^b$}}\fi\chkspace}

\def\pafbzo{\relax\ifmmode \tilde{A}_{FB}|_{O(0)} \else
{{$\tilde{A}_{FB}|_{O(0)}$}}\fi\chkspace}
\def\pafbfo{\relax\ifmmode \tilde{A}_{FB}|_{\oalp} \else
{{$\tilde{A}_{FB}|_{\oalp}$}}\fi\chkspace}
\def\pafbso{\relax\ifmmode \tilde{A}_{FB}|_{\oalpsq} \else
{{$\tilde{A}_{FB}|_{\oalpsq}$}}\fi\chkspace}
\def\pafbto{\relax\ifmmode \tilde{A}_{FB}|_{\oalpc} \else
{{$\tilde{A}_{FB}|_{\oalpc}$}}\fi\chkspace}

\def\pafbbzo{\relax\ifmmode \tilde{A}_{FB}^b|_{O(0)} \else
{{$\tilde{A}_{FB}^b|_{O(0)}$}}\fi\chkspace}
\def\pafbbfo{\relax\ifmmode \tilde{A}_{FB}^b|_{\oalp} \else
{{$\tilde{A}_{FB}^b|_{\oalp}$}}\fi\chkspace}
\def\pafbbso{\relax\ifmmode \tilde{A}_{FB}^b|_{\oalpsq} \else
{{$\tilde{A}_{FB}^b|_{\oalpsq}$}}\fi\chkspace}
\def\pafbbto{\relax\ifmmode \tilde{A}_{FB}^b|_{\oalpc} \else
{{$\tilde{A}_{FB}^b|_{\oalpc}$}}\fi\chkspace}

\def\afbo0{\tilde{A}_{FB}|_{O(0)}}
\def\afbo1{\tilde{A}_{FB}|_{\oalp}}
\def\afbo2{\tilde{A}_{FB}|_{\oalpsq}}
\def\afbo3{\tilde{A}_{FB}|_{\oalpc}}

\def\lam{\relax\ifmmode \Lambda_{\overline{MS}}
       \else {{$\Lambda_{\overline{MS}}$}}\fi\chkspace}
\def\lamuds{\relax\ifmmode \Lambda^{(3)}_{\overline{MS}}
       \else {{$\Lambda^{(3)}_{\overline{MS}}$}}\fi\chkspace}
\def\lamudsc{\relax\ifmmode \Lambda^{(4)}_{\overline{MS}}
       \else $\Lambda^{(4)}_{\overline{MS}}$\fi\chkspace}
\def\lamudscb{\relax\ifmmode \Lambda^{(5)}_{\overline{MS}}
       \else $\Lambda^{(5)}_{\overline{MS}}$\fi\chkspace}

\def\alp{\relax\ifmmode \alpha_s\else $\alpha_s$\fi\chkspace}
\def\alpbar{\relax\ifmmode \bar{\alpha_s}
       \else $\bar{\alpha_s}$\fi\chkspace}
\def\alpmz{\relax\ifmmode \alpha_s(M_Z)\else $\alpha_s(M_Z)$\fi\chkspace}
\def\alpmzsq{\relax\ifmmode \alpha_s(M_Z^2)
       \else $\alpha_s(M_Z^2)$\fi\chkspace}

\def\oalp{\relax\ifmmode O(\alpha_s)\else{{O($\alpha_s$)}}\fi\chkspace}
\def\oalpsq{\relax\ifmmode O(\alpha_s^2)
           \else{{O($\alpha_s^2$)}}\fi\chkspace}
\def\oalpc{\relax\ifmmode O(\alpha_s^3)
           \else{{O($\alpha_s^3$)}}\fi\chkspace}
\def\oalpf{\relax\ifmmode O(\alpha_s^4)
           \else{{O($\alpha_s^4$)}}\fi\chkspace}

\def\rb{\relax\ifmmode R_3^b/R_3^{all}
           \else{{$R_3^b/R_3^{all}$}}\fi\chkspace}
\def\rc{\relax\ifmmode R_3^c/R_3^{all}
           \else{{$R_3^c/R_3^{all}$}}\fi\chkspace}
\def\ruds{\relax\ifmmode R_3^{uds}/R_3^{all}
           \else{{$R_3^{uds}/R_3^{all}$}}\fi\chkspace}
\def\ri{\relax\ifmmode R_3^i/R_3^{all}
           \else{{$R_3^i/R_3^{all}$}}\fi\chkspace}
\def\rj{\relax\ifmmode R_3^j/R_3^{all}
           \else{{$R_3^j/R_3^{all}$}}\fi\chkspace}
\def\alpi{\relax\ifmmode \alpha^i_s/\alpha^{all}_s
           \else{{$\alpha^i_s/\alpha^{all}_s$}}\fi\chkspace}

\def\mbz{\relax\ifmmode m_b(M_Z)
           \else{{$m_b(M_Z)$}}\fi\chkspace}
\def\mbb{\relax\ifmmode m_b(M_b)
           \else{{$m_b(M_b)$}}\fi\chkspace}

\def\z0{\relax\ifmmode Z^0 \else {$Z^0$} \fi\chkspace}
\def\h0{\relax\ifmmode H^0 \else {$H^0$} \fi\chkspace}
\def\Dst{\relax\ifmmode {\rm D}^* \else {D$^*$}\fi\chkspace}
\def\Dpl{\relax\ifmmode {\rm D}^+ \else {D$^+$}\fi\chkspace}
\def\D0{\relax\ifmmode {\rm D}^0 \else {D$^0$}\fi\chkspace}
\def\Kst{\relax\ifmmode {\rm K}^* \else {K$^*$}\fi\chkspace}
\def\K0{\relax\ifmmode {\rm K}^0_s \else {K$^0_s$}\fi\chkspace}
\def\Kpl{\relax\ifmmode {\rm K}^+ \else {K$^+$}\fi\chkspace}
\def\Kstz{\relax\ifmmode {\rm K}^{*0} \else {K$^{*0}$}\fi\chkspace}

\def\beq{\begin{equation}}
\def\eeq{\end{equation}}
\def\bea{\begin{eqnarray}}
\def\eea{\end{eqnarray}}

\renewcommand{\baselinestretch}{1.5}

 1

\catcode`\@=11 
\makeatletter
\def\@seccntformat#1{\csname the#1\endcsname.\hskip 1em}

\makeatother
\begin{document}
\thispagestyle{empty}
\begin{flushright}
{\footnotesize\renewcommand{\baselinestretch}{.75}
  SLAC--PUB--8503\\
July 2000\\
}
\end{flushright}

\vskip 1truecm
 
\begin{center}
\Large \bf 

{First Symmetry Tests in}

{Polarized \z0 Decays to \bbg}
 
\vspace {1.0cm}
 
\end{center}

\normalsize
 
\begin{center}
 {\bf The SLD Collaboration$^*$}\\
Stanford Linear Accelerator Center \\
Stanford University, Stanford, CA~94309
\end{center}
 
\vspace{1cm}
 
\begin{center}
{\bf ABSTRACT }
\end{center}

{\small 
\noindent
We have made the first direct symmetry tests in the 
decays of polarized \z0 bosons into fully-identified
\bbg states, collected in the SLD experiment at SLAC.
We searched for evidence of parity violation at the \bbg vertex by studying the 
asymmetries in the $b$-quark 
polar- and azimuthal-angle distributions, and for evidence of 
T-odd, CP-even or odd, final-state interactions by measuring
angular correlations between the three-jet plane and the \z0 polarization.
We found results consistent with Standard Model expectations and set
95\% C.L. limits on anomalous contributions.
}

\centerline{({\it Submitted to Physical Review Letters})}

\vfill
\noindent
Contributed to the XXX International Conference on High Energy Physics,
July 27--August 2 2000, Osaka, Japan; Ref.~689.

{\footnotesize
\noindent
Work supported by Department of Energy contract DE-AC03-76SF00515 (SLAC).}

\eject

Polarization is an essential tool in investigations of fundamental symmetries
in elementary-particle interactions. Parity violation
was first discovered using $\beta$ decays of polarized $^{60}$Co~\cite{beta}, 
and T, CP and
CPT violations were searched for using polarized neutrons~\cite{neutron}
and polarized positronium~\cite{posi}.
The unique sample of polarized \z0 bosons produced in annihilations of 
longitudinally-polarized electrons with unpolarized positrons at the SLAC 
Linear Collider (SLC) can similarly be employed for fundamental symmetry tests.
Here we use \ep \ra \z0 \ra
three-jet events to test symmetry properties of the Standard Model (SM).
The \bbg final state provides a
particularly interesting probe for possible beyond-SM
processes that couple to massive particles.

The tree-level differential cross section for \ep $\rightarrow$
\qqg can be expressed~\cite{qqg}
$$2\pi{{d^4\sigma}\over{d(\cos\theta)d\chi dx d\overline{x}}} = 
[~{3\over 8}(1+\cos^2\theta)\,\sigma_U+
{3\over4}\sin^2\theta\,\sigma_L 
$$
\begin{equation}
+{3\over4}\sin^2\theta\cos 2\chi\,\sigma_T
+{3\over{2\sqrt{2}}}\sin 2\theta\cos\chi\,\sigma_I]~h_f^{(1)} 
+~[~{3\over4}\cos\theta\, \sigma_P
-{3\over\sqrt{2}}\sin\theta\cos\chi\,\sigma_A~]~
h_f^{(2)},
\label{eqsigma}
\end{equation}
where $\theta$ is the polar angle of the thrust axis w.r.t. the electron beam,
and $\chi$ is the azimuthal angle of the event plane w.r.t. the quark-electron
plane. 
The thrust axis is defined to be along the most energetic jet and to
point into (opposite) the hemisphere containing the quark (antiquark)
if the quark (antiquark) has the higher energy. The sign of cos$\chi$ is defined
in terms of momentum vectors, 
sign(cos$\chi$) = sign(($\vec{q}\times\vec{g})\cdot(\vec{q}\times\vec{e^-}$)).
The functions $h_f^{(1)}$ and $h_f^{(2)}$ contain the dependence on the
beam polarization and the electroweak couplings~\cite{qqg};
QCD contributions are expressed in terms of
$\sigma_i$ $\equiv$ $d^2\sigma_i/dx d\overline{x}$, with
$i$ = $U,L,T,I,A,P$, where
$x$ and $\bar x$ are the scaled momenta of the quark and anti-quark,
respectively, 
While the first four terms are even under P reversal, 
the last two terms are P-odd, and are sensitive to any parity-violating
interactions at the $Z^0{\rm q}{\bar {\rm q}}$ or \qqg vertices. 
More generally there can be,
in addition,  three terms that are odd 
under T reversal~\cite{Hagiwara}. However, these terms 
vanish at tree level in a theory that respects CPT invariance.

We report the first experimental 
study of angular asymmetries in polarized \z0 decays to fully-identified
\bbg states. We present new tests of QCD using the P-odd asymmetries in the 
$\theta$ and $\chi$
distributions~\cite{Phil}, as well as the first measurements of two T-odd
triple-product correlations between the \bbg plane and the \z0 polarization.
Integrating Eq.~\ref{eqsigma} over $x$, $\overline{x}$ and $\chi$, 
\begin{equation}
{{d\sigma} \over {d\cos\theta}} \propto 1\,+\,\alpha
\cos^2\theta\,+\,2 P_Z\; A_P \;\cos\theta, 
\end{equation}
where $P_Z$ = $(P_{e^-}-A_e)/(1-P_{e^-}\cdot A_e)$, 
$P_{e^-}$ is the signed electron-beam polarization, 
the parity-violation parameter
$A_P = A_b\cdot\hat{\sigma}_P/(\hat{\sigma}_U+\hat{\sigma}_L)$,
$A_e ~(A_b)$
is the electroweak coupling of the $Z^0$ to the initial 
(final) state, and $A_j = 2v_ja_j/(v_j^2+a_j^2)$ in terms of
the vector, $v_j$, and axial-vector, $a_j$,  couplings of fermion $j$ to the \z0; 
$\alpha =
(\hat{\sigma}_U-2\hat{\sigma}_L)/(\hat{\sigma}_U+2\hat{\sigma}_L)$, and
 $\hat{\sigma}_i \equiv \int{d^2\sigma_i}$.   
Similarly, 
\begin{equation}
{{d\sigma}\over {d\chi}} \propto \,1+\,\beta
\cos 2\chi \,- \,{3\pi\over {2\sqrt{2}}}\;P_Z\; A_P' \,\cos\chi, 
\end{equation}
where $A_P' = A_b\cdot \hat{\sigma}_A/(\hat{\sigma}_U+\hat{\sigma}_L)$,
and $\beta = \hat{\sigma}_T/(\hat{\sigma}_U+\hat{\sigma}_L)$.
Given the value of $A_b$, measurement of $A_P$ and $A_P'$ allows one to test the 
QCD predictions for $\hat{\sigma}_P/(\hat{\sigma}_U+\hat{\sigma}_L)$ and 
 $\hat{\sigma}_A/(\hat{\sigma}_U+\hat{\sigma}_L)$. Furthermore, the ratio
$A_P'/A_P$ yields $\hat{\sigma}_A/\hat{\sigma}_P$ independently of $A_b$. 
  
In terms of the polar angle $\omega$, w.r.t. the electron-beam direction, 
of the vector ${\vec n}$ normal to the event plane,
\begin{equation}
{{d\sigma} \over {d\cos\omega}} \propto 1\,+\,\gamma 
\cos^2\omega\,+\,{16 \over 9} \;P_Z\;A_T  \cos\omega, 
\end{equation}
where 
$\gamma = (2\hat{\sigma}_L-\hat{\sigma}_U-6\hat{\sigma}_T)/
(3\hat{\sigma}_U+2\hat{\sigma}_L+2\hat{\sigma}_T)$;
the asymmetry term is one of the three T-odd terms mentioned above.
The vector $\vec{n}$ can be defined in several ways; for example: 1)
using the two highest-energy jets,
${\vec n} = \vec{p}_1 \times \vec{p}_2$; 
or 2)  using the quark and anti-quark
momenta, ${\vec n} = \vec{p}_b \times \vec{p}_{\bar b}$. 
The asymmetry 
is CP-even in the first definition, and CP-odd in the second;
in both cases in the SM $A_T$ = 0 at 
tree level. Higher-order corrections to
\ep$\rightarrow$\bbg yield $|A_T| < 10^{-5}$~\cite{Brandenburg}.
The asymmetry in $\cos\omega$ is hence potentially sensitive to
beyond-SM processes~\cite{Murayama}.   
  
The measurement was performed 
using approximately 550,000 \z0 \ra hadrons decays 
produced in collisions of longitudinally-polarized electrons with 
unpolarized positrons at the SLC between 1993 and 1998. 
In order to reduce systematic effects on polarization-dependent asymmetries
the electron polarization direction was reversed randomly pulse-by-pulse; 
the magnitude of the average polarization was 0.73.
The data were recorded in the SLC Large Detector (SLD)~\cite{sld}.
The trigger and hadronic event selection criteria are described 
elsewhere~\cite{trig}.
This analysis used charged tracks measured in the
central drift chamber 
and in the CCD-based vertex detectors (VXD)~\cite{vxd}. 
About 70\% of the data were taken with the new VXD installed
in 1996 (VXD3), and the rest with the previous detector (VXD2).
Only well-reconstructed 
tracks~\cite{wellrecon} were used for the $b$-jet tagging.
Particle energies were measured in the liquid-argon and warm-iron 
calorimeters; the calorimetric information was 
used only to reconstruct the thrust axis.
  
In each event jets were
reconstructed using the ``Durham''  algorithm~\cite{durham}.
To select planar three-jet events we required 
exactly three reconstructed jets to be found with a
jet-resolution parameter value of
$y_c$=0.005, the sum of the angles
between the three jets to be greater than 358$^\circ$,
and that each jet contain at least two charged tracks;
74,886 events satisfied these criteria. The jet energies were
calculated by using the measured jet directions and solving the
three-body kinematics assuming massless jets.
The jets were then labeled such that $E_1 > E_2 > E_3$.

To select \bbg events the long lifetime and large invariant mass of 
$B$-hadrons were exploited.
An algorithm ~\cite{zvtop,sldbfrag} was applied to the set of
well reconstructed tracks in each jet in an attempt to reconstruct a decay vertex. 
Vertices were required to contain at least two tracks, and to 
be separated from the interaction point (IP) by at least 1 mm.
We calculated (described below) that
the probability for reconstructing at least one such vertex was
$\sim$ 91\% (77\%) in \bbg events, $\sim$ 45\% (26\%) in \ccg events, and
$\sim$~2\% (2\%) in light-quark events recorded in VXD3 (VXD2).
 
An event was selected as \bbg if at least one jet contained a vertex with 
invariant mass~\cite{sldbfrag} $>$ 1.5 GeV/c$^2$. 
A total of 14,658 events satisfied 
this requirement and were subjected to further analysis.
We calculated that this selection is 84\% (69\%) efficient for 
identifying a sample of \bbg events with 84\% (87\%) purity, and 
containing 14\% (11\%)
\ccg and 2\% (2\%) light-flavor backgrounds.     

In order to isolate the gluon jet,
a $b$-tag was evaluated for each jet in the selected event sample.
This tag was defined to be positive if the jet contained a
vertex, or if it contained at
least 3 ``significant'' tracks, \ie with normalized  impact
parameter w.r.t the IP $d/\sigma_d >$ 3~\cite{wellrecon}; 
otherwise the tag was defined to be negative. 
Jet 1 was tagged as the gluon jet ($g$) if it had a negative
$b$-tag and both jets 2 and 3 had positive $b$-tags.
Jet 2 was tagged
as the gluon-jet if it had a negative $b$-tag and jet 3 a positive
one. Otherwise, jet 3 was tagged as the gluon jet. We calculated that,
on average, the purity of the tagged gluon jet sample was 91\% (88\%).

In order to distinguish the $b$ and $\bar{b}$ jets we evaluated
for each jet $j$ the momentum-weighted charge,
$ Q_j = \sum q_i |\vec{p}_i\cdot \vec{\hat{t}}|^\kappa$, 
where $\kappa$=0.5, $\vec{\hat{t}}$ is the unit vector along the
thrust axis and $q_i$ and $\vec{p}_i$ are the
charge and momentum of the $i^{th}$ track in jet $j$.
We calculated the charge difference, 
$Q_{diff}$ $\equiv$ $Q_1 - Q_2 - Q_3$. If $Q_{diff}$ 
was negative (positive) we assigned jet 1 as the 
$b$ ($\bar{b}$) jet, unless jet 1 was gluon tagged, in which case
we assigned jet 2 as the  $b$ ($\bar{b}$) jet
if $Q_{diff}$ was positive (negative). 
The probability of correct $b$/$\overline{b}$ assignment, $\cal{P}(Q_{diff})$, was 
calculated from the data by using a self-calibration technique~\cite{junk}, 
$\cal{P}(Q_{diff}) = 1 / (1+e^{-\alpha_b |Q_{diff}|})$, 
where $\alpha_b$ is a fitted parameter. 
Averaged over $\cos\theta$ $\alpha_b$ = 0.218$\pm$0.018 (VXD3) (0.248$\pm$0.030 (VXD2)),
corresponding to $<\cal{P}>$ = 0.68 (VXD3) (0.67 (VXD2)).
 
A JETSET 7.4~\cite{JETSET}-based Monte Carlo simulation of hadronic \z0
decays combined with a simulation of the detector
response was used to evaluate the efficiency and purity of the \bbg
event selection and the purity of the jet flavor tags.
For those simulated events satisfying the three-jet criteria,
exactly three jets were
reconstructed at the parton level by applying the jet algorithm to
the parton four-momenta. The
three parton-level jets were associated with the three detector-level
jets by choosing the
combination that minimized the sum of the angular differences between the
 corresponding jet axes, and the energies and charges of the matching
jets were compared.

Figs. 1(a) and (b) show the observed $\cos\theta$ distributions 
for event samples produced by 
left- and right-handed electron beams, respectively.
The distributions may be described by
\begin{eqnarray}
\lefteqn{{{d\sigma} \over {d\cos\theta}} \propto 1 + \alpha 
\cos^2\theta~ + 2 P_Z \,\cos\theta~[ A_P\,f_b\,(2\,p^b_{\theta}-1) + }
\nonumber\\
& &  \qquad A_{P,c}\,f_c\,(2\,p^c_{\theta}-1) + A_{P,uds}\,(1-f_b-f_c)\,
(2\,p^{uds}_{\theta}-1)~], 
\label{diffb}
\end{eqnarray}
where $f_b$ and $f_c$ are the fractions of \bbg and \ccg events in the sample, 
respectively, $A_{P,c}$ and $A_{P,uds}$ are the 
asymmetry parameters for the respective backgrounds (Fig.~1),  
and $p^b_{\theta}$, $p^c_{\theta}$ and $p^{uds}_{\theta}$ are the
probabilities to sign $\cos\theta$ correctly in \bbg, \ccg and  
light-quark events, respectively. 
The data were used to calculate $p^b_{\theta}$ ( = $\cal{P}$ above); 
all other quantities were calculated from the simulation.                                 
A  maximum-likelihood
fit of Eq.~\ref{diffb} yielded $A_P = 0.855 \pm 0.050$ (stat.).

Figs. 2(a) and (b) show the observed $\chi$ distributions for events
produced with left- and right-handed electron beams, respectively.
They may be described by
\begin{eqnarray}
\lefteqn{{{d\sigma} \over {d\chi}}\, \propto \,1 \,+ \,\beta 
\cos 2\chi\, - 
 {3\pi \over{2\sqrt{2}}}\,P_Z\,\cos\chi~
[ A_P'\,f_b\,(2p^b_{\chi}-1) + }
\nonumber\\
& &  \qquad A_{P,c}'\,f_c\,(2p^c_{\chi}-1) + A_{P,uds}'\,(1-f_b-f_c)\,(2p^{uds}_{\chi}-1)~], 
\label{diffc}
\end{eqnarray}
where $A_{P,c}'$ and $A_{P,uds}'$ are the 
asymmetry parameters for the respective backgrounds, and 
$p^b_{\chi}$, $p^c_{\chi}$ and $p^{uds}_{\chi}$ are the
probabilities to sign $\cos\chi$ correctly in \bbg, \ccg and  
light-quark events, respectively.
Averaged over $\chi$, $p^b_{\chi}$ = 0.64 (0.63), which
was derived using the measured value of $\cal{P}$ (above) combined with
the simulated probability to tag the gluon jet correctly.
All other parameters were calculated from the simulation.                                 
A  maximum-likelihood
fit of Eq.~\ref{diffc} yielded $A_P' = -0.013 \pm 0.033$ (stat.).

Figs. 3(a) and (b) show the left-right-forward-backward asymmetry in 
$|\cos\omega|$ $\equiv$ $z$,
$$\tilde{A}_{FB}(z) \equiv {{\sigma_L(z)-\sigma_L(-z)
+ \sigma_R(-z)-\sigma_R(z) } \over
{\sigma_L(z)+\sigma_L(-z)
+\sigma_R(-z)+\sigma_R(z)}}
$$  
for the two definitions of the direction of $\vec{n}$:
(1) $\vec{p_1}\times\vec{p_2}$, and (2) $\vec{p_b}\times\vec{p_{\bar b}}$. 
No asymmetry is apparent. The cos$\omega$ distributions may be described,
assuming no asymmetries in the \ccg and light-quark backgrounds, by
\begin{equation}
\label{t-odd-fit}
{{d\sigma} \over {d\cos\omega}} \propto 1 - {1\over3} 
\cos^2\omega+{16 \over 9}P_Z\,A^{\pm}_T  f_b (2p^b_{\pm}-1)\cos\omega, 
\label{diffd}
\end{equation}
where $p^b_{\pm}$ is the
probability to sign $\cos\omega$ correctly in the CP-even (+) and CP-odd ($-$)
cases, respectively.
In the CP-even case, in which 
the jets were labeled according to their energy,
six detector-jet energy orderings were possible for a given parton-jet
energy ordering. 
Using the simulation we calculated that, 
averaged over cos$\omega$, $p^b_+$ = 0.76 (0.76).
In the CP-odd case, both the gluon jet and the
$b$-jet must be tagged correctly, and $p^b_-$ ( = $p^b_{\chi}$ above) 
= 0.64 (0.63). 
Maximum-likelihood fits (Fig.~3) of Eq.~\ref{t-odd-fit} yielded
$A_T^+ = -0.014 \pm 0.016$ (stat.) and
$A_T^- = -0.035 \pm 0.024$ (stat.).

We considered sources of systematic error~\cite{junk,rbprl} which potentially affect
our results. The modeling of the detector response
was studied by varying the tracking efficiency and resolution
within their estimated uncertainties, and by varying the 
vertex-mass and significant-track requirements.
In addition, the error on the probability for correct jet-energy ordering was
estimated from the difference between results derived using
HERWIG~\cite{HERWIG} and JETSET.
The error on the $b$/$\overline{b}$-jet identification probability  
was derived from the statistical uncertainty in the 
$\alpha_b$ determination using the self-calibration technique.
Contributions from the modeling of underlying physics processes were also
studied. In \bb events we considered the uncertainties
on: the branching fraction for \z0$\rightarrow$\bb, the $B$-hadron 
fragmentation function, the rates of production
of $B^\pm$, $B^0$ and $B^0_s$ mesons, and $B$ baryons, the lifetimes of $B$
mesons and baryons, and the average $B$-hadron decay charge multiplicity.
In \cc events we considered the uncertainties on: the branching
fraction for \z0$\rightarrow$\cc, the charmed hadron fragmentation
function, the rates of production of $D^0$, $D^+$ and $D_s$ mesons, and
charmed baryons, and the charged multiplicity of charmed hadron decays.  
We also considered the rate of production of $s\bar{s}$ in the jet
fragmentation process, and the production of secondary
\bb and \cc from gluon splitting.
The uncertainty on the beam polarization was also taken into
account.

The systematic errors on $A_P'$ and $A_T^{\pm}$ are negligibly small as the
uncertainty diminishes with the measured asymmetry, which in all three cases
is small and consistent with zero.
In the case of $A_P$ the largest error contributions 
arose from the $\alpha_b$ uncertainty ($\delta A_P/A_P \sim$5.7\%) and from
the uncertainty in the background level ($\delta A_P/A_P \sim$3.9\%); other
contributions were at the level of 1\% or less. 

From the measured values of $A_P$ and $A_P'$, and 
assuming the SM expectation 
of $A_b$ $\simeq$ 0.935 for $\sin^2\theta_w$=0.232, we find, respectively
$${
{\hat{\sigma}_P} \over {\hat{\sigma}_U+\hat{\sigma}_L}} 
= 0.914 \pm 0.053\, (stat.)\pm 0.063\, (syst.),
$$
$${{\hat{\sigma}_A} \over {\hat{\sigma}_U+\hat{\sigma}_L}} 
= -0.014 \pm 0.035\,(stat.) \pm 0.002 \,(syst.).
$$
These values are consistent with the ${\cal O}(\alpha_s^2)$ QCD expectation 
(for massless quarks) of
$\hat{\sigma}_P/(\hat{\sigma}_U+\hat{\sigma}_L)$ = 0.93 and 
$\hat{\sigma}_A/(\hat{\sigma}_U+\hat{\sigma}_L)$ = $-0.06$, respectively,
calculated using JETSET 7.4.
These yield $\hat{\sigma_A}/\hat{\sigma_P}=-0.015\pm0.038$ 
independent of the assumed $A_b$, and consistent
with the expected value $-0.065$.
We also used the $A_P$ and $A_P'$ values to set a 95\% C.L. 
limit on an anomalous axial-vector coupling of the gluon to the 
$b$-quark, parametrized~\cite{ogreid} 
by a factor $(1+\epsilon\gamma_5)\gamma_{\nu}$
in the \bbg coupling, of $\epsilon$ $<$ 0.34.
The measured values of $A_T^+$ and $A_T^-$ correspond to    
95\% C.L. limits of $-0.045 < A_T^+ < 0.016$ and $-0.082 < A_T^- < 0.012$
(Fig.~3).

In conclusion, we have made the first symmetry tests in
polarized \z0 decays to \bbg. 
From the forward-backward polar-angle asymmetry of the signed-thrust axis
and the azimuthal-angle asymmetry we found the parity-violation 
parameters $A_P$ and $A_P'$, respectively, 
to be consistent with ${\cal O}(\alpha_s^2)$ QCD expectations. We set a 
corresponding 95\% C.L. limit on parity violation at the \bbg vertex.
Using the event-plane normal polar-angle distributions
we set 95\% C.L. limits on the T-odd, CP-even and odd
asymmetry parameters $A_T^+$ and $A_T^-$, respectively. 
 
We thank Arnd Brandenburg, Lance Dixon and Per Osland for many helpful discussions.
We thank the personnel of the SLAC accelerator department and the
technical
staffs of our collaborating institutions for their outstanding efforts
on our behalf.

\vskip .5truecm

\vbox{\footnotesize\renewcommand{\baselinestretch}{1}\noindent
$^*$Work supported by Department of Energy
  contracts:
  DE-FG02-91ER40676 (BU),
  DE-FG03-91ER40618 (UCSB),
  DE-FG03-92ER40689 (UCSC),
  DE-FG03-93ER40788 (CSU),
  DE-FG02-91ER40672 (Colorado),
  DE-FG02-91ER40677 (Illinois),
  DE-AC03-76SF00098 (LBL),
  DE-FG02-92ER40715 (Massachusetts),
  DE-FC02-94ER40818 (MIT),
  DE-FG03-96ER40969 (Oregon),
  DE-AC03-76SF00515 (SLAC),
  DE-FG05-91ER40627 (Tennessee),
  DE-FG02-95ER40896 (Wisconsin),
  DE-FG02-92ER40704 (Yale);
  National Science Foundation grants:
  PHY-91-13428 (UCSC),
  PHY-89-21320 (Columbia),
  PHY-92-04239 (Cincinnati),
  PHY-95-10439 (Rutgers),
  PHY-88-19316 (Vanderbilt),
  PHY-92-03212 (Washington);
  The UK Particle Physics and Astronomy Research Council
  (Brunel, Oxford and RAL);
  The Istituto Nazionale di Fisica Nucleare of Italy
  (Bologna, Ferrara, Frascati, Pisa, Padova, Perugia);
  The Japan-US Cooperative Research Project on High Energy Physics
  (Nagoya, Tohoku);
  The Korea Research Foundation (Soongsil, 1997).}

\vfill
\eject

\noindent
{\bf $^{*}$List of authors}

%
%
%
\begin{center}
\def\iAOMORI{$^{(1)}$}
\def\iBRI{$^{(2)}$}
\def\iBRUN{$^{(3)}$}
\def\iBU{$^{(4)}$}
\def\iCOLO{$^{(5)}$}
\def\iCSU{$^{(6)}$}
\def\iFERR{$^{(7)}$}
\def\iFRAS{$^{(8)}$}
\def\iJHU{$^{(9)}$}
\def\iLBL{$^{(10)}$}
\def\iMASS{$^{(11)}$}
\def\iMISSI{$^{(12)}$}
\def\iMIT{$^{(13)}$}
\def\iMOSCOW{$^{(14)}$}
\def\iNAGO{$^{(15)}$}
\def\iOREG{$^{(16)}$}
\def\iOXF{$^{(17)}$}
\def\iPERU{$^{(18)}$}
\def\iRAL{$^{(19)}$}
\def\iRUTG{$^{(20)}$}
\def\iSLAC{$^{(21)}$}
\def\iSOONG{$^{(22)}$}
\def\iTENN{$^{(23)}$}
\def\iTOHO{$^{(24)}$}
\def\iUCSB{$^{(25)}$}
\def\iUCSC{$^{(26)}$}
\def\iVAND{$^{(27)}$}
\def\iWASH{$^{(28)}$}
\def\iWISC{$^{(29)}$}
\def\iYALE{$^{(30)}$}

  \baselineskip=.75\baselineskip
\mbox{Koya Abe\unskip,\iTOHO}
\mbox{Kenji Abe\unskip,\iNAGO}
\mbox{T. Abe\unskip,\iSLAC}
\mbox{I. Adam\unskip,\iSLAC}
\mbox{H. Akimoto\unskip,\iSLAC}
\mbox{D. Aston\unskip,\iSLAC}
\mbox{K.G. Baird\unskip,\iMASS}
\mbox{C. Baltay\unskip,\iYALE}
\mbox{H.R. Band\unskip,\iWISC}
\mbox{T.L. Barklow\unskip,\iSLAC}
\mbox{J.M. Bauer\unskip,\iMISSI}
\mbox{G. Bellodi\unskip,\iOXF}
\mbox{R. Berger\unskip,\iSLAC}
\mbox{G. Blaylock\unskip,\iMASS}
\mbox{J.R. Bogart\unskip,\iSLAC}
\mbox{G.R. Bower\unskip,\iSLAC}
\mbox{J.E. Brau\unskip,\iOREG}
\mbox{M. Breidenbach\unskip,\iSLAC}
\mbox{W.M. Bugg\unskip,\iTENN}
\mbox{D. Burke\unskip,\iSLAC}
\mbox{T.H. Burnett\unskip,\iWASH}
\mbox{P.N. Burrows\unskip,\iOXF}
\mbox{A. Calcaterra\unskip,\iFRAS}
\mbox{R. Cassell\unskip,\iSLAC}
\mbox{A. Chou\unskip,\iSLAC}
\mbox{H.O. Cohn\unskip,\iTENN}
\mbox{J.A. Coller\unskip,\iBU}
\mbox{M.R. Convery\unskip,\iSLAC}
\mbox{V. Cook\unskip,\iWASH}
\mbox{R.F. Cowan\unskip,\iMIT}
\mbox{G. Crawford\unskip,\iSLAC}
\mbox{C.J.S. Damerell\unskip,\iRAL}
\mbox{M. Daoudi\unskip,\iSLAC}
\mbox{S. Dasu\unskip,\iWISC}
\mbox{N. de Groot\unskip,\iBRI}
\mbox{R. de Sangro\unskip,\iFRAS}
\mbox{D.N. Dong\unskip,\iMIT}
\mbox{M. Doser\unskip,\iSLAC}
\mbox{R. Dubois\unskip,}
\mbox{I. Erofeeva\unskip,\iMOSCOW}
\mbox{V. Eschenburg\unskip,\iMISSI}
\mbox{E. Etzion\unskip,\iWISC}
\mbox{S. Fahey\unskip,\iCOLO}
\mbox{D. Falciai\unskip,\iFRAS}
\mbox{J.P. Fernandez\unskip,\iUCSC}
\mbox{K. Flood\unskip,\iMASS}
\mbox{R. Frey\unskip,\iOREG}
\mbox{E.L. Hart\unskip,\iTENN}
\mbox{K. Hasuko\unskip,\iTOHO}
\mbox{S.S. Hertzbach\unskip,\iMASS}
\mbox{M.E. Huffer\unskip,\iSLAC}
\mbox{X. Huynh\unskip,\iSLAC}
\mbox{M. Iwasaki\unskip,\iOREG}
\mbox{D.J. Jackson\unskip,\iRAL}
\mbox{P. Jacques\unskip,\iRUTG}
\mbox{J.A. Jaros\unskip,\iSLAC}
\mbox{Z.Y. Jiang\unskip,\iSLAC}
\mbox{A.S. Johnson\unskip,\iSLAC}
\mbox{J.R. Johnson\unskip,\iWISC}
\mbox{R. Kajikawa\unskip,\iNAGO}
\mbox{M. Kalelkar\unskip,\iRUTG}
\mbox{H.J. Kang\unskip,\iRUTG}
\mbox{R.R. Kofler\unskip,\iMASS}
\mbox{R.S. Kroeger\unskip,\iMISSI}
\mbox{M. Langston\unskip,\iOREG}
\mbox{D.W.G. Leith\unskip,\iSLAC}
\mbox{V. Lia\unskip,\iMIT}
\mbox{C. Lin\unskip,\iMASS}
\mbox{G. Mancinelli\unskip,\iRUTG}
\mbox{S. Manly\unskip,\iYALE}
\mbox{G. Mantovani\unskip,\iPERU}
\mbox{T.W. Markiewicz\unskip,\iSLAC}
\mbox{T. Maruyama\unskip,\iSLAC}
\mbox{A.K. McKemey\unskip,\iBRUN}
\mbox{R. Messner\unskip,\iSLAC}
\mbox{K.C. Moffeit\unskip,\iSLAC}
\mbox{T.B. Moore\unskip,\iYALE}
\mbox{M. Morii\unskip,\iSLAC}
\mbox{D. Muller\unskip,\iSLAC}
\mbox{V. Murzin\unskip,\iMOSCOW}
\mbox{S. Narita\unskip,\iTOHO}
\mbox{U. Nauenberg\unskip,\iCOLO}
\mbox{H. Neal\unskip,\iYALE}
\mbox{G. Nesom\unskip,\iOXF}
\mbox{N. Oishi\unskip,\iNAGO}
\mbox{D. Onoprienko\unskip,\iTENN}
\mbox{L.S. Osborne\unskip,\iMIT}
\mbox{R.S. Panvini\unskip,\iVAND}
\mbox{C.H. Park\unskip,\iSOONG}
\mbox{I. Peruzzi\unskip,\iFRAS}
\mbox{M. Piccolo\unskip,\iFRAS}
\mbox{L. Piemontese\unskip,\iFERR}
\mbox{R.J. Plano\unskip,\iRUTG}
\mbox{R. Prepost\unskip,\iWISC}
\mbox{C.Y. Prescott\unskip,\iSLAC}
\mbox{B.N. Ratcliff\unskip,\iSLAC}
\mbox{J. Reidy\unskip,\iMISSI}
\mbox{P.L. Reinertsen\unskip,\iUCSC}
\mbox{L.S. Rochester\unskip,\iSLAC}
\mbox{P.C. Rowson\unskip,\iSLAC}
\mbox{J.J. Russell\unskip,\iSLAC}
\mbox{O.H. Saxton\unskip,\iSLAC}
\mbox{T. Schalk\unskip,\iUCSC}
\mbox{B.A. Schumm\unskip,\iUCSC}
\mbox{J. Schwiening\unskip,\iSLAC}
\mbox{V.V. Serbo\unskip,\iSLAC}
\mbox{G. Shapiro\unskip,\iLBL}
\mbox{N.B. Sinev\unskip,\iOREG}
\mbox{J.A. Snyder\unskip,\iYALE}
\mbox{H. Staengle\unskip,\iCSU}
\mbox{A. Stahl\unskip,\iSLAC}
\mbox{P. Stamer\unskip,\iRUTG}
\mbox{H. Steiner\unskip,\iLBL}
\mbox{D. Su\unskip,\iSLAC}
\mbox{F. Suekane\unskip,\iTOHO}
\mbox{A. Sugiyama\unskip,\iNAGO}
\mbox{A. Suzuki\unskip,\iNAGO}
\mbox{M. Swartz\unskip,\iJHU}
\mbox{F.E. Taylor\unskip,\iMIT}
\mbox{J. Thom\unskip,\iSLAC}
\mbox{E. Torrence\unskip,\iMIT}
\mbox{T. Usher\unskip,\iSLAC}
\mbox{J. Va'vra\unskip,\iSLAC}
\mbox{R. Verdier\unskip,\iMIT}
\mbox{D.L. Wagner\unskip,\iCOLO}
\mbox{A.P. Waite\unskip,\iSLAC}
\mbox{S. Walston\unskip,\iOREG}
\mbox{A.W. Weidemann\unskip,\iTENN}
\mbox{E.R. Weiss\unskip,\iWASH}
\mbox{J.S. Whitaker\unskip,\iBU}
\mbox{S.H. Williams\unskip,\iSLAC}
\mbox{S. Willocq\unskip,\iMASS}
\mbox{R.J. Wilson\unskip,\iCSU}
\mbox{W.J. Wisniewski\unskip,\iSLAC}
\mbox{J.L. Wittlin\unskip,\iMASS}
\mbox{M. Woods\unskip,\iSLAC}
\mbox{T.R. Wright\unskip,\iWISC}
\mbox{R.K. Yamamoto\unskip,\iMIT}
\mbox{J. Yashima\unskip,\iTOHO}
\mbox{S.J. Yellin\unskip,\iUCSB}
\mbox{C.C. Young\unskip,\iSLAC}
\mbox{H. Yuta\unskip.\iAOMORI}

\it
  \vskip \baselineskip                   
  \baselineskip=.75\baselineskip   
\iAOMORI
  Aomori University, Aomori, 030 Japan, \break
\iBRI
  University of Bristol, Bristol, United Kingdom, \break
\iBRUN
  Brunel University, Uxbridge, Middlesex, UB8 3PH United Kingdom, \break
\iBU
  Boston University, Boston, Massachusetts 02215, \break
\iCOLO
  University of Colorado, Boulder, Colorado 80309, \break
\iCSU
  Colorado State University, Ft. Collins, Colorado 80523, \break
\iFERR
  INFN Sezione di Ferrara and Universita di Ferrara, I-44100 Ferrara, Italy,
\break
\iFRAS
  INFN Laboratori Nazionali di Frascati, I-00044 Frascati, Italy, \break
\iJHU
  Johns Hopkins University,  Baltimore, Maryland 21218-2686, \break
\iLBL
  Lawrence Berkeley Laboratory, University of California, Berkeley, California
94720, \break
\iMASS
  University of Massachusetts, Amherst, Massachusetts 01003, \break
\iMISSI
  University of Mississippi, University, Mississippi 38677, \break
\iMIT
  Massachusetts Institute of Technology, Cambridge, Massachusetts 02139, \break
\iMOSCOW
  Institute of Nuclear Physics, Moscow State University, 119899 Moscow, Russia,
\break
\iNAGO
  Nagoya University, Chikusa-ku, Nagoya, 464 Japan, \break
\iOREG
  University of Oregon, Eugene, Oregon 97403, \break
\iOXF
  Oxford University, Oxford, OX1 3RH, United Kingdom, \break
\iPERU
  INFN Sezione di Perugia and Universita di Perugia, I-06100 Perugia, Italy,
\break
\iRAL
  Rutherford Appleton Laboratory, Chilton, Didcot, Oxon OX11 0QX United Kingdom,
\break
\iRUTG
  Rutgers University, Piscataway, New Jersey 08855, \break
\iSLAC
  Stanford Linear Accelerator Center, Stanford University, Stanford, California
94309, \break
\iSOONG
  Soongsil University, Seoul, Korea 156-743, \break
\iTENN
  University of Tennessee, Knoxville, Tennessee 37996, \break
\iTOHO
  Tohoku University, Sendai, 980 Japan, \break
\iUCSB
  University of California at Santa Barbara, Santa Barbara, California 93106,
\break
\iUCSC
  University of California at Santa Cruz, Santa Cruz, California 95064, \break
\iVAND
  Vanderbilt University, Nashville,Tennessee 37235, \break
\iWASH
  University of Washington, Seattle, Washington 98105, \break
\iWISC
  University of Wisconsin, Madison,Wisconsin 53706, \break
\iYALE
  Yale University, New Haven, Connecticut 06511. \break

\rm
%

\end{center}

\vfill
\eject

\epsfbox{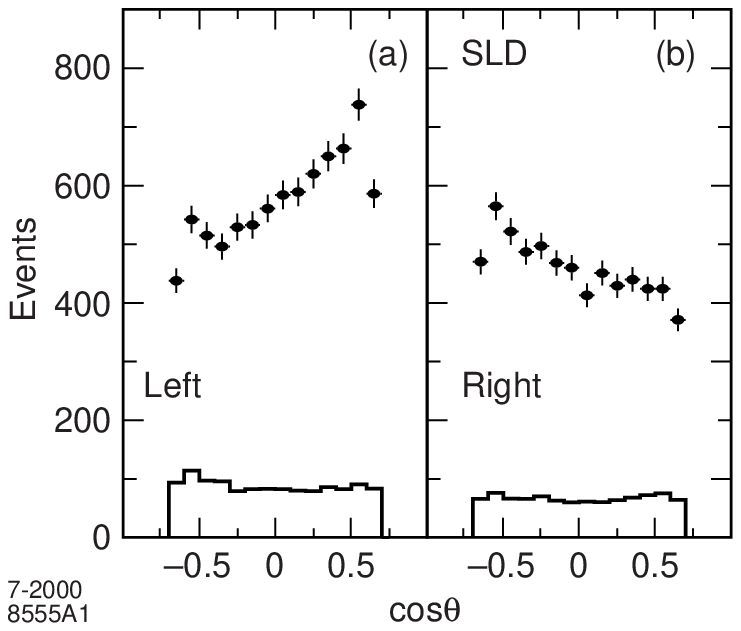}

\noindent
{\bf Figure 1}.
Polar-angle distribution of the signed-thrust axis direction (see text)
for (a) left- and (b) right-handed electron beam.
The histograms represent the simulated backgrounds.

\vskip\baselineskip 

\vskip 1truecm

\epsfbox{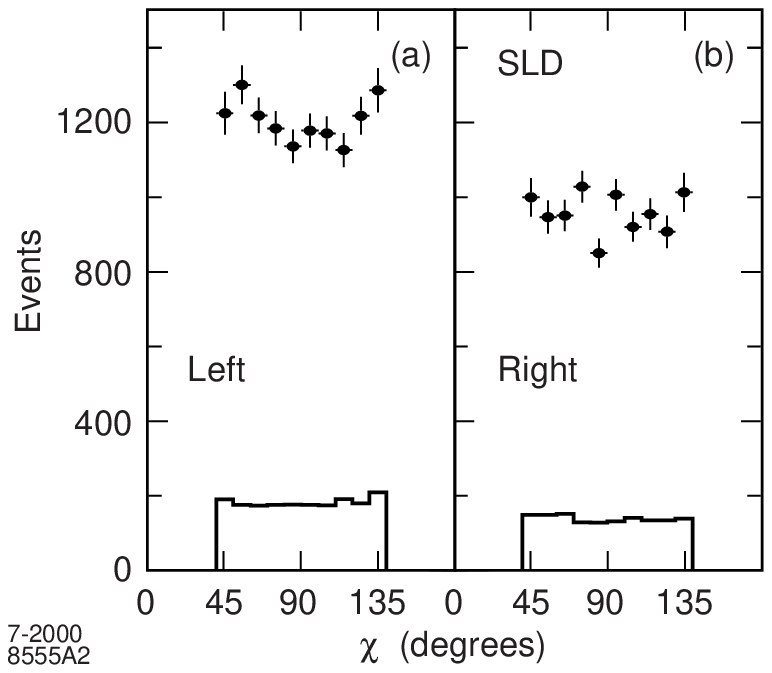}

\noindent
{\bf Figure 2}.
As Fig.~1, for the azimuthal-angle distribution (see text).

\vskip\baselineskip 

\vskip 1truecm

\epsfbox{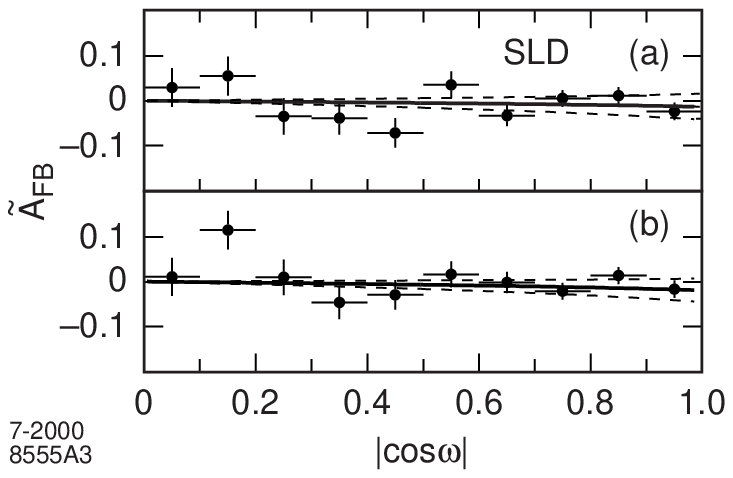}

\noindent
{\bf Figure 3}.
Left-right forward-backward asymmetry in $\cos\omega$ (see text)
for (a) CP-even case, and (b) CP-odd case.
The solid curve is the best fit to the data sample, and the dashed curves 
correspond to the 95\% C.L. limits. 

\end{document}